\documentclass[aps,prl,reprint,groupedaddress]{revtex4}

\usepackage{bbm}
\usepackage{graphicx}
\usepackage{subfigure}
\usepackage{amsmath}
\begin{document}

\title{Nonthermal entanglement dynamics in a dipole-facilitated glassy model with disconnected subspaces}

\author{Guanhua Chen$^{1}$ and Yao Yao$^{1,2}$\footnote{Electronic address:~\url{yaoyao2016@scut.edu.cn}}}

\address{$^1$ Department of Physics, South China University of Technology, Guangzhou 510640, China\\
$^2$ State Key Laboratory of Luminescent Materials and Devices, South China University of Technology, Guangzhou 510640, China}

\date{\today}

\begin{abstract}
We construct a dipole-facilitated kinetic constraint to partition the Hilbert space into three disconnected subspaces, two of which are nonthermal and the other acts as an intrinsic thermal bath. The resulting glassy system freely oscillates in nonthermal subspaces, making the quantum entanglement perform like a substantial qubit. The spatially spreading entanglement, quantified by concurrence, fidelity and 2-R\'{e}nyi entropy, is found to be spontaneously recovered which is absent in other reference models. Under low-frequency random flip noise, this reversible hydrodynamics of entanglement holds high fidelity and volume law, while at high frequency thermalization unusually occurs leading to a strange phase transition. Our work offers an elaborate space structure for realizing ergodicity breaking and controllable entanglement dynamics.
\end{abstract}

\maketitle

In traditional thermodynamic perspective, the entropy simply increases when more states in the Hilbert space are populated. Recent progresses in random quantum circuits however indicate the measurement can induce breakdown of entropic volume law, as a unitary makes the spreading entanglement entropy increase by one in each step, implying its particle nature \cite{1,2}. It is then pretty interesting whether the entanglement can be manipulated as a substantial particle, but just like in superconducing circuits \cite{3,4}, the uniform levels of particle hinder efficient quantum control and it has to be reduced to two-level system enabled by specific anharmonicity. Subsequently, it serves as the main object of the present work to introduce well-designed kinetic constraints to disconnect the subspaces to which the entanglement may spread.

In glassy systems, kinetic constraints give rise to weak ergodic breaking in presence of the fragmentation of Hilbert space with outlying nonthermal states  \cite{5,6,7}. For example, in a triangular antiferromagnetic lattice, the dynamics of a local spin is frustrated due to the constraint from neighboring spins \cite{8,9}. In the spin glass phase, therefore, long-range orders are absent and the featured timescale of thermalization becomes ultraslow. This interesting local dynamics is then described by kinetically constrained models (KCMs), such as the Fredrickson-Andersen (FA) model and the East model \cite{10,11,12}. Based on the disorder-free quantum East model, it has been proven that a large number of nonthermal states can be constructed to manifest area-law entanglement entropy \cite{13,14}. In this Letter, we introduce a new dipole-facilitated kinetic constraint, which results in unusual ergodicity breaking by two disconnected subspaces. Initially from the Bell state and the Greenberger-Horne-Zeilinger (GHZ) state, we will study the exotic dynamics of entanglement within this elaborately constructed block structure of Hilbert space.

\begin{figure}[htbp]
	\includegraphics[scale=1]{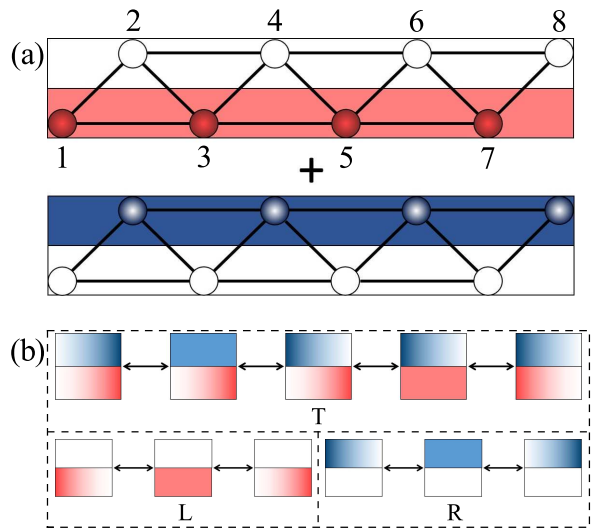}
	\caption{\label{fig1} (a) Schematic of an 8-site GHZ state with the circle representing that site is in spin-down state and the ball being spin-up. Red and blue zone denote the state of relevant sites are changeable. (b) A thermal subspace (T) and two nonthermal subspaces (L and R) are figured out in disconnected block representation. The upper and lower halves in each square (solid: $\uparrow\uparrow$, white: $\downarrow\downarrow$, gradient: $\uparrow\downarrow$ or $\downarrow\uparrow$) represent even and odd sites, respectively.  }
\end{figure}

Our basic idea is to constrain the spin flip by some specific neighboring state, which is motivated by the CNOT gate in quantum computation with the control circuit being instead composed of two spins or a dipole. We start from cutting a triangular spin lattice into a quasi-one-dimensional chain as sketched in Fig.~\ref{fig1}(a) \cite{15,16}. Odd and even sites are separated into lower and upper sides labeled by different colors. The constraint rule is described as follows. First, two neighboring spins favor antiparallel configuration due to antiferromagnetic interaction. Second, a third spin in the triangle formed with these two spins is allowed to flip only if its neighboring spin is in down state. Of course we can parallel consider the constraint from up state but that is surely equivalent. Subsequently, this newly-proposed KCM can be named as dipole-facilitated model (DFM). The model Hamiltonian is written as
\begin{equation}\label{eq1}
	\begin{aligned}
		H_{\rm df}=\sum_{i=1}^{L} (Q_i P_{i+1}X_{i+2}+X_i P_{i+1}Q_{i+2}),
	\end{aligned}
\end{equation}
where $X_i$ is the x-Pauli operator on $i$-th site, $Q_i=|\uparrow\rangle \langle \uparrow|_i$ and $P_i=|\downarrow\rangle \langle \downarrow|_i$ are projectors of spin-up and spin-down, respectively. Throughout this work, the periodic boundary condition (PBC) is adopted for this model.

On the potential experimental realization of the model, we first notice that the Rydberg atoms, an ideal experimental platform for KCMs, have been used to study nonequilibrium and slow dynamics \cite{17,18,19}. The excited and ground state of atom can be mapped as spin up and down. Under some controllable experimental conditions, the nearest-neighbour Rydberg blockade effect can be well described by the so-called PXP model \cite{20,21}, which has got very similar constraint with the present DFM. In this context, we think that the DFM should not be difficult to be realized in a Rydberg atom array.

Let us first discuss the Hilbert space of DFM in block representation \cite{22}. We consider four sites as instance. Two successive sites are grouped into one ``group-site", and we denote ($\downarrow\downarrow$), ($\uparrow\downarrow$), ($\downarrow$$\uparrow$) and ($\uparrow\uparrow$) as $\circ$, $\triangleright$, $\triangleleft$ and $\bullet$, respectively. Under the action of $H_{\rm df}$, we notice that these configurations ($\circ\circ$), ($\triangleleft\triangleright$), ($\triangleleft\bullet$), ($\bullet\triangleright$) and ($\bullet\bullet$) are annihilated and the rest can be categorized into three subspaces. The transformation rules are
\begin{equation}\label{eq2}
	\begin{aligned}
		&{\rm T}:\circ\bullet\longleftrightarrow \triangleright\bullet \longleftrightarrow \triangleright\triangleleft \longleftrightarrow \bullet\triangleleft \longleftrightarrow \bullet\circ, \\
        {\rm L}:\circ\triangleright&\longleftrightarrow \triangleright\triangleright\longleftrightarrow \triangleright\circ, \quad {\rm R}:\circ\triangleleft\longleftrightarrow \triangleleft\triangleleft\longleftrightarrow\triangleleft\circ. \\
	\end{aligned}
\end{equation}
These three reaction paths can be straightforwardly extended to any longer chains and subsequently construct a large thermal subspace labeled by T and two nonthermal subspaces labeled by L (all even sites are spin-down) and R (all odd sites are spin-down), as sketched in Fig.\ref{fig1}(b).

Most interestingly, these three disconnected subspaces totally decide the active spatial state of systems. For example, if the initial state of a system without perturbation is $|\uparrow\downarrow\dots\uparrow\downarrow\rangle$, even sites will persistently stay in spin-down state, implying that the time evolution of the system is confined to the subspace L. The reverse holds true as well, namely states initiated from $|\downarrow\uparrow\dots\downarrow\uparrow\rangle$ stays in the R subspace. Now, if we have an initial state $|0\rangle=|\downarrow\downarrow\dots\downarrow\downarrow\rangle$, it will evolve into completely different subspace and stay there depending on how we add spin-up into the lattice. Adding a single spin-up to even (odd) site activates R (L) subspace, respectively, and two successive spin-up's break the confinement of above two nonthermal subspaces leading to the T subspace which may act as an intrinsic thermal bath as discussed below.

One may ask if these disconnected subspaces possess similar feature with that in the Hilbert space fragmentation. Breaking the Hilbert space into disconnected sectors is the nontrivial feature of generic KCM, but in most cases the number of Krylov subspaces is exponentially dependent of the system size \cite{19}. In our DFM the number of subspaces is fixed no matter how large the system is, which is as stated essential to produce nontrivial entanglement dynamics.

\begin{figure*}[htbp]
	\includegraphics[scale=1]{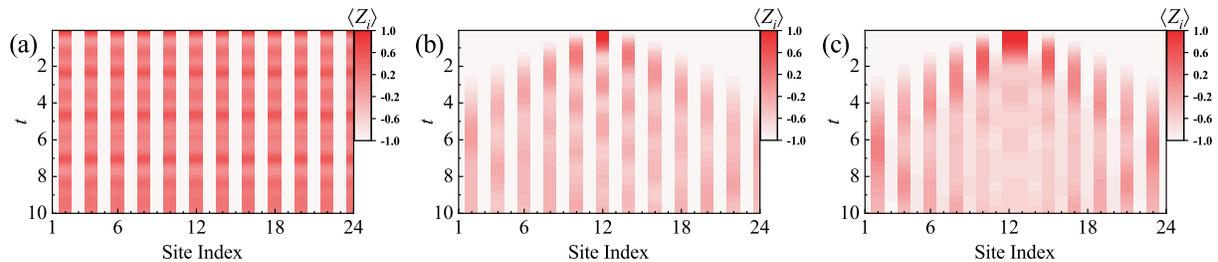}
	\caption{\label{fig2} Time evolution of $\langle Z_i \rangle$ on a 24-site lattice. (a) The initial state is $|\downarrow\uparrow\dots\downarrow\uparrow\rangle$. The odd sites keep in spin-down and even sites behave periodic oscillation. (b) The initial state is $|\downarrow\dots\downarrow\uparrow\downarrow\dots\downarrow\rangle$, i.e. a single spin-up is located at $i=12$, which spreads to the ends of the chain but odd sites keep in spin-down persistently. (c) The initial state is $|\downarrow\dots\downarrow\uparrow\uparrow\downarrow\dots\downarrow\rangle$, i.e. the $i=12$ and $13$ sites are in spin-up. Differently, spin-up states will cover all sites after sufficiently long time.}
\end{figure*}

In the following, we use time-evolving block decimation (TEBD) to compute the dynamics on the chain \cite{23,24}. We first show the time evolution of expectation value of z-Pauli operator $\langle Z_i\rangle$ with different initial states for $L=24$. Fig.~\ref{fig2}(a) and (b) show the results from the initial states $|\downarrow\uparrow\dots\downarrow\uparrow\rangle$ and $|\downarrow\dots\downarrow\uparrow\downarrow\dots\downarrow\rangle$, respectively. In two cases, the spins are continuously flipped on the even sites, but all odd sites keep spin-down without any flipping. That is, the system freely oscillates in the R subspace resulting from ergodicity breaking. The parallel case in L subspace is not shown. For a comparison, we calculate the initial state with two successive spin-up sites in the middle of the chain, namely $|\downarrow\dots\downarrow\uparrow\uparrow\downarrow\dots\downarrow\rangle$ as shown in Fig.\ref{fig2}(c). This will transform the subspace L and R to T and the system becomes ergodic, as all sites are being flipped in the time evolution.

Let us now categorize all sites into two for convenience with regard to quantum computations. The sites always being spin-down in a product state are called idle sites and the others as work sites. As long as the system is sufficiently large, the probability that all work sites simultaneously flip to spin-down is extremely low. Therefore, we can denote idle sites to be 0, work sites together as 1, to form a large and substantial logical qubit. In superconducting circuits, quantum non-demolition parity measurements can be realized by assigning some qubits as control nodes to detect errors of adjacent data qubits \cite{25}. Analogously, herein, if we introduce a spin-up at a left end work site, the right end work site will tell us the initial odevity, and errors caused by noise can be detected by the appearance of successive spin-up's. A single logical qubit can thus be in a superposition state of product states in the L and R subspace, labeled by $|L\rangle$ or $|R\rangle$ respectively, paving a novel way for fault tolerance with this ``big" qubit.

\begin{figure*}[htbp]
	\includegraphics[scale=1.1]{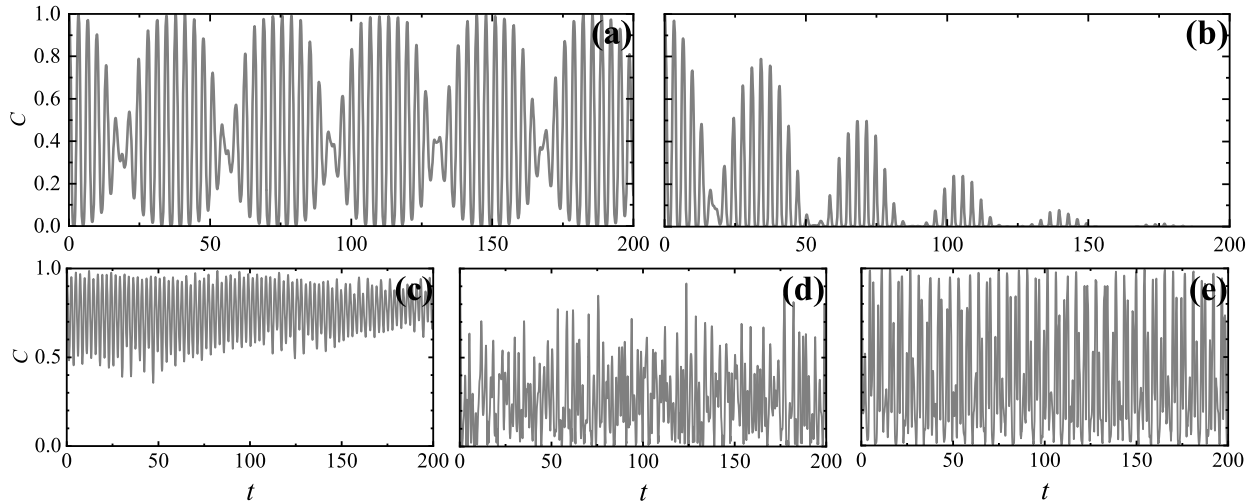}
\caption{\label{fig3} Concurrence between middle two sites ($i=4$ and 5) in 8-site systems. All initial states are set as the Bell state. (a) Time evolution with DFM shows recoverable periodic oscillations. (b) Adding a random flip noise with $T_X=1$, the bipartite entanglement gradually decreases during oscillation. This calculation result is averaged over 1000 samples for random $X$ operators. Results of time evolution without noise under (c) the East model at Rokhsar-Kivelson point, (d) the PXP model from Rydberg blockade and (e) the antiferromagnetic Heisenberg model do not show periodic behavior. The East model is in open boundary condition and others are in PBC.}
\end{figure*}

Essentially, this logical qubit can be promisingly applied to various entangled states. By a two-qubit gate, it is easy to obtain a local Bell state in many-body system ($L=8$), i.e., $|\rm Bell\rangle=\frac{1}{\sqrt{2}}(|\downarrow\downarrow\downarrow\downarrow\uparrow\downarrow\downarrow\downarrow\rangle+|\downarrow\downarrow\downarrow\uparrow\downarrow\downarrow\downarrow\downarrow\rangle)$, which will generally be evolving into $|L\rangle+|R\rangle$. Then, we focus on the entanglement between middle two sites to see the influence of the disconnected subspaces. To this end, we calculate the concurrence to quantize the bipartite entanglement between them \cite{26,27}, which is defined as
\begin{equation}\label{eq4}
	C(\rho)=\rm max\{\lambda_1-\lambda_2-\lambda_3-\lambda_4,0\},
\end{equation}
where $\lambda_k$'s are the eigenvalues in descending order of the Hermitian matrix $r=\sqrt{\sqrt{\rho}\tilde{\rho}\sqrt{\rho}}$ and $\tilde{\rho}=(\sigma_{\rm y} \otimes \sigma_{\rm y})\rho^*(\sigma_{\rm y} \otimes \sigma_{\rm y})$. $\sigma_{\rm y}$ and $\rho$ are y-Pauli matrix and reduced density matrix of the two middle sites by partially tracing others \cite{28}.

As displayed in Fig.~\ref{fig3}(a), we observe a surprising periodicity of the entanglement, especially its maximum, which is very like a Newton's cradle. This novel nonthermal dynamic behavior stems from that two superposed bases of $|\rm Bell\rangle$ are restricted in two disconnected nontrivial subspaces, and the time evolution in these two subspaces are completely symmetric allowing the entanglement to be spontaneously recovered after spreading. To be comparisons, three other typical models of spin chain are also calculated, namely the East model, the PXP model and the antiferromagnetic Heisenberg model as shown in Fig.~\ref{fig3}(c), (d) and (e). There are no conserved quantities in the PXP model which was first introduced in the quantum many-body scars and also in the classical East model for describing spin glasses \cite{14}. The Heisenberg model merely conserves the total spin. Remarkably, concurrences in these three models do not perform any visible cradle-like periodicity but just irregular oscillations.

To see how the nonthermal dynamics is broken down, we then mimic the inevitable noise which can be described as (off-diagonal) bit flip and the (diagonal) projection, respectively. In the first case if an idle site is flipped from spin-down to spin-up by off-diagonal noise, one of subspaces L or R will be transferred into T. As a result, the entangled state is destroyed irreversibly by the T thermal bath. We then add a random measurement into the Hamiltonian as a noisy source, following the form:
\begin{equation}\label{eq6}
	H_{\rm X}(t)=H_{\rm df}+\sum_{n} X_j \delta(t-nT_{\rm X})
\end{equation}
where $j$ is a random site to be measured and the noisy term occurs every $t=nT_{\rm X}$. In Fig.~\ref{fig3}(b), we observe that although the concurrence still oscillates, the envelope decays gradually under the noise $T_{\rm X}=1$ and will not recover, manifesting the irreversible breaking of periodicity. It implies the measurement has induced a transition of the dynamic feature. In the case of diagonal noise, we can also set a noisy Hamiltonian similar to Eq.~(\ref{eq6}) with projectors $Q_i$ or $P_i$. Compared with the off-diagonal noise which connects subspaces, these diagonal projectors do not hybridize the subspaces unless all work sites are simultaneously set to be 0 which is almost impossible, so their influence is much easier to be eliminated and is not considered here.

\begin{figure}[htbp]
	\includegraphics[scale=1]{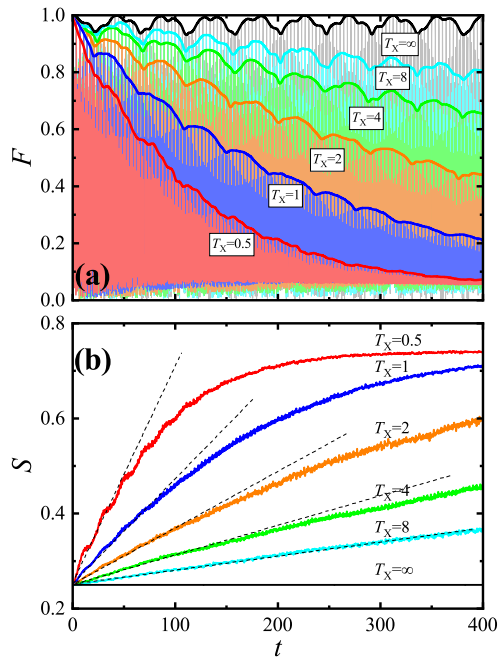}
	\caption{\label{fig4} Fidelity and reduced 2-R\'{e}nyi entropy evolving with $H_{\rm X}$. (a) The fidelity $F=|\langle{\rm GHZ}| \phi(t)\rangle| $ with $T_{\rm X}=0.5$ (red), $T_{\rm X}=1$ (blue), $T_{\rm X}=2$ (orange), $T_{\rm X}=4$ (green), $T_{\rm X}=8$ (cyan) and $T_{\rm X}=\infty$ (black) is obtained after quench from 8-site $|\rm GHZ\rangle$. The system behaves high-frequency oscillations and the envelope lines of higher bound are bolded. The $T_{\rm X}=\infty$ case is equal to evolving under $H_{\rm df}$, which shows periodicity without decay. (b) The reduced 2-R\'{e}nyi entropy $S(\rho)$ of even sites for different $T_{\rm X}$. At $T_{\rm X}=\infty$, the entropy is constant over time. Dashed lines indicate the linear dependency of entropy and time, namely the volume law. Above calculation results are all averaged over 500 runs for $X$ operators on random sites.}
\end{figure}

In order to further comprehend the measurement-induced phase transition in DFM, we move on to consider the 8-site GHZ state $|\rm GHZ\rangle=\frac{1}{\sqrt{2}}(|\uparrow\downarrow\uparrow\downarrow\uparrow\downarrow\uparrow\downarrow\rangle+|\downarrow\uparrow\downarrow\uparrow\downarrow\uparrow\downarrow\uparrow\rangle)$ as the initial state which also belongs to $|L\rangle+|R\rangle$ and has many-body entangled properties distinguished from the Bell state. We calculate the fidelity $F=|\langle{\rm GHZ}| \phi(t)\rangle|$ between evolving and GHZ state to characterize the influence of wrong actions under noise, which actually quantifies the fluctuation of initial state. As shown in Fig.~\ref{fig4}(a), for the system without random actions, the fidelity is able to come back to $1$ periodically, implying perfect many-body revivals. By adding a noisy term, the fidelity is gradually decreasing over time, i.e. the distance between evolving state and $|{\rm GHZ}\rangle$ is increasing. If setting the time axis in the unit of nanosecond as for usual case, it is estimated that even in a very noisy environment with frequency being about $125$ MHz, the fidelity of $|\rm GHZ\rangle$ can keep above $0.8$ after $400$ ns, suggesting a fairly good robustness \cite{29}.

We can also calculate the entanglement entropy between even and odd sites. The reduced 2-R\'{e}nyi entropy with a normalized logarithmic base for the reduced density matrix $\rho$ of even sites is then defined as
\begin{equation}\label{eq9}
	S(\rho)=-\log_{16} \rm Tr(\rho^2),
\end{equation}
which also characterizes the localization of wave function \cite{30}. As depicted in Fig.~\ref{fig4}(b), without noise the entanglement remains a fixed value $S_{\rm min}=\log_{16}2=0.25$, with respect to two nonzero diagonal elements in the reduced density matrix. In other cases, the R\'{e}nyi entropy keeps growing before saturated. There is a significant phase transition: For $T_{\rm X}>1$ the entropy is in the early stage linearly dependent of time suggesting the system stays in an entangling phase with volume law, and for $T_{\rm X}<1$ the entropy rapidly saturates indicating the measurement induces an area-law disentangling phase and thermalization takes place instead of localization in common cases. More interestingly, the area-law entropy is even larger than the volume-law entropy. In the small nonthermal subspaces the entanglement is periodic and localized so the increase is extremely slow, while when the localization is broken down, a large number of states will instantly evolve into large thermal T subspace to make the entropy rapidly increase, which is very similar with the entanglement tsunami \cite{31}. Consequently, the T subspace itself performs an intrinsic thermal bath for nonthermal subspaces which is the most appealing structure of DFM.

Before ending, we make a simple analogy between our model and the disorder-free localization model which is composed of spinless fermions and spin-1/2 on the bond \cite{32}. As a fermionic matter field, the fermion tunneling is related to coupled spins. We take one triangle on the lattice as instance. It is easy to derive that
\begin{equation}\label{eq10}
	\begin{aligned}
		H_{\rm df}^2&=[Q_1P_2(\sigma^+_3+\sigma^-_3)+(\sigma^+_1+\sigma^-_1)P_2Q_3]^2 \\
		&=Q_1P_2+P_2 Q_3+\sigma^-_1 P_2\sigma^+_3+\sigma^+_1 P_2\sigma^-_3,\end{aligned}
\end{equation}
where $\sigma^+_i=|\uparrow\rangle\langle\downarrow|$ and $\sigma^-_i=|\downarrow\rangle\langle\uparrow|$ are spin upper and lower operators. One can see that, the last two terms are nothing but the constrained hopping term in the disorder-free localization model, namely the hopping is determined by spin on bond. As a result, our model $H_{\rm df}$ just decomposes the localization into two subspaces and thus enables the nonthermal entanglement.

In summary, originated from triangular frustrated kinetic constraints, the DFM manifests exotic disconnected subspace structure. The entanglement exhibits appealing nonthermal periodic dynamics acting as a substantial qubit. Under low-frequency flip noise, the entanglement ballistically increases while at high frequency it saturates in an instant, indicating a phase transition to thermalization takes place stemming from the intrinsic thermal subspace.

\section*{Acknowledgment}

The authors gratefully acknowledge support from the Special Project for Research and Development in Key Areas of
Guangdong Province (Grant No.~2020B0303300001), and National Natural Science Foundation of China (Grant Nos.~11974118).

\bibliography{dfm_v27.bbl}
\end{document}